\documentclass[english]{emulateapj}
\usepackage[latin9]{inputenc}
\usepackage{graphicx}
\usepackage{amsmath}

\def\degree{\ifmmode {^\circ}\else {$^\circ$}\fi}
\def\mum{\ifmmode {\rm \mu {\rm m}}\else $\rm \mu {\rm m}$\fi}
\def\arcsec{\ifmmode ^{\prime \prime}\else $^{\prime \prime}$\fi}

\def\inch{\ifmmode ^{\prime \prime}\else $^{\prime \prime}$\fi}
\def\arcmin{\ifmmode ^{\prime}\else $^{\prime}$\fi}

\def\mjup{\ifmmode {\rm M_J}\else $\rm M_J$\fi}
\def\rjup{\ifmmode {\rm R_J}\else $\rm R_J$\fi}
\def\mearth{\ifmmode {\rm M_{\oplus}}\else $\rm M_{\oplus}$\fi}
\def\rearth{\ifmmode {\rm R_{\oplus}}\else $\rm R_{\oplus}$\fi}
\def\lsun{\ifmmode {\rm L_{\odot}}\else $\rm L_{\odot}$\fi}
\def\msun{\ifmmode {\rm M_{\odot}}\else $\rm M_{\odot}$\fi}
\def\mjupyr{\ifmmode {\rm M_J~yr^{-1}}\else $\rm M_J~yr^{-1}$\fi}
\def\msunyr{\ifmmode {\rm M_{\odot}~yr^{-1}}\else $\rm M_{\odot}~yr^{-1}$\fi}
\def\kms{\ifmmode {\rm km~s^{-1}}\else $\rm km~s^{-1}$\fi}

\def\gcm3{g~cm$^{-3}$}

\makeatletter

\providecommand{\tabularnewline}{\\}

\makeatother

\makeatother

\usepackage{babel}

\begin{document}

\title{Wind-accretion disks in wide binaries, \\
second generation protoplanetary disks and accretion onto white
dwarfs}

\author{Hagai B. Perets$^{1}$ and Scott J. Kenyon$^{2}$}

\email{hperets@physics.technion.ac.il}

\affil{$^{1}$Deloro fellow, Technion - Israel Institute of Technology, Haifa, Israel\\
$^{2}$Harvard-Smithsonian Center for Astrophysics, 60 Garden St.,Cambridge, MA, USA 02138}
\begin{abstract}
Mass transfer from an evolved donor star to its binary companion is
a standard feature of stellar evolution in binaries.  In wide binaries,
the companion star captures some of the mass ejected in a wind by the 
primary star. The captured material forms an accretion disk. Here, we study 
the evolution of wind-accretion disks, using a numerical approach which 
allows us to follow the long term evolution. For a broad range of initial 
conditions, we derive the radial density and temperature profiles of the 
disk. In most cases, wind-accretion leads to long-lived stable disks over 
the lifetime of the AGB donor star. The disks have masses of a few times 
$10^{-5}-10^{-3}$ $M_{\odot}$, with surface density and temperature
profiles that follow broken power-laws. The total mass in the disk 
scales approximately linearly with the viscosity parameter used. 
Roughly 50\% to 80\% of the
mass falling into the disk accretes onto the central star; the rest 
flows out through the outer edge of the disk into the stellar wind 
of the primary. For systems with large accretion rates, the secondary
accretes as much as 0.1 \msun. When the secondary is a white dwarf,
accretion naturally leads to nova and supernova eruptions. For all types 
of secondary star, the surface density and temperature profiles of massive
disks resemble structures observed in protoplanetary disks, suggesting 
that coordinated observational programs might improve our understanding
of uncertain disk physics.
\end{abstract}

\section{Introduction}

Mass transfer from an evolved donor star to a companion star is a
common outcome of stellar evolution in binary systems. In close binaries
with periods of 1--2 yr or less, the donor can fill its tidal surface
and transfer mass rapidly into a massive disk surrounding the companion
\citep[e.g.,][]{ken+82,bat+82b,ken+91,siv+09}. Although these disks are luminous,
the systems are short-lived and rarely observable \citep{ken+84, ken86}.

Disks formed via wind accretion may be more common. Chemically peculiar
stars, including Barium stars and CH stars, are probably polluted
by material accreted from a binary companion \citep{mcc+90,luc+91,han+95,bus+01}.
White dwarfs (WDs) accreting material through disks filled by the
winds of their evolved companions are often identified as symbiotic
stars or very slow (also symbiotic) novae \citep{ken+83,ken86,web+87,sok+06}.
In both cases, long-lived disks are possible when the donor star does
not fill its tidal lobe.

Theoretical studies of the physical properties of wind-accretion disks
provide an important framework for interpreting observations of these
systems. The replenishment of newly accreted material and the formation
of a disk may also have important implications for the formation and
evolution of planetary systems. A newly formed disk may interact with
pre-existing planets, leading to the re-growth and/or migration of
these planets, or possibly the formation new planets and/or debris
disks \citep{per10b,per10a}.

Several previous studies explored the formation of wind-accretion
disks through SPH and AMR simulations \citep{mas+98,dev+09}. Here we use
a different simplified analytical/numerical approach typically used
for the study of disks in shorter period binaries and in protostellar/protoplanetary
systems. This approach allows us to follow the long term evolution
of the disks, which is more difficult in hydro simulations, and to compare 
the derived physical structure with protoplanetary disks. We focus on 
the accretion from low mass companions in wide binaries ($3-100\,$ AU),
which evolve through wind-accretion.

We begin with a description of the basic conditions in wind-accreting
binaries, followed by a detailed explanation of the methods we use
to follow their evolution. We then describe results of the wind-accretion
disk structure followed by a discussion and summary.

\section{Wind accretion in wide binaries}

To understand the conditions required for the formation of circumstellar
disks from wind accreted material, we follow \citet{sok+00}. For
a disk to form, $j_{a}$, the specific angular momentum of the accreted
material, must exceed $j_{2}=(GM_{2}R_{2})^{1/2}$, the specific angular
momentum of a particle in a Keplerian orbit at the equator of the
accreting star of radius $R_{2}$. For accretion from a wind, the
net specific angular momentum of the material entering the Bondi-Hoyle
accretion radius, $R_{a}$, is $j_{BH}=0.5(2\pi/P)R_{a}^{2}$ \citep{wan81},
where $P$ is the orbital period.  The actual accreted specific angular
momentum for high Mach number flows is $j_{a}=\eta j_{BH}$, where 
$\eta\sim0.1$ and $\eta\sim0.3$ for isothermal and adiabatic flows,
respectively \citep{liv+86}.
For a binary composed of a main sequence (MS) accretor with mass $M_{2}$
and a mass-losing asymptotic giant branch (AGB) star with mass $M_{1}$,
requiring $j_{a}>j_{2}$ leads to a simple condition for disk formation
\begin{align}
1<\frac{j_{a}}{j_{2}} \simeq  1.2\left(\frac{\eta}{0.2}\right)\left(\frac{M_{1}+M_{2}}{2.5M_{\odot}}\right)\left(\frac{M_{2}}{M_{\odot}}\right)^{3/2}\nonumber \\
 \times\left(\frac{R_{2}}{R_{\odot}}\right)^{-1/2}\left(\frac{a}{100AU}\right)^{-3/2}\left(\frac{v_{r}}{10km\, s^{-1}}\right)^{-4}\label{eq:mass-accretion}\end{align}
 where $R_{2}$ is the radius of the accreting star, $a$ is the semi-major
axis of the binary, and $v_{r}$ is the relative velocity of the wind
and the accretor. 

Adopting $R_{2}\sim R_{\odot}$, disks can form in binaries with $a\lesssim$
10--100 AU for $M_{1}\approx$ 1--10 \msun\ and $v_{r}\approx~5-20$~\kms.
This range in $a$ spans the peak of the distribution of binary star
separations \citep{duq+91}. Thus, many binaries contain wind-accretion
disks during the late stages of their evolution.

Red giant branch (RGB) and AGB stars lose mass in low velocity winds
at typical rates of $\dot{M}_{w}\sim10^{-8}-10^{-5}$ \msunyr\ 
\citep[e.g.,][]{taysea84,ken86,olo+02,ram+06}.
To estimate the fraction of this material accreted by the MS companion,
we adopt the Bondi-Hoyle rate for an isotropic wind from the mass-losing 
star, \begin{equation}
\dot{M}_{a}=\left(\frac{R_{a}}{2a}\right)^{2}~\dot{M}_{w}~,\label{eq:mdot_bondi}\end{equation}
 where $R_{a}$ is the Bondi-Hoyle accretion radius \begin{equation}
R_{a}=\frac{2GM_{2}}{v_{r}^{2}+c_{s}^{2}}~.\label{eq:r_bondi}\end{equation}
 In this expression, $v_{r}^{2}=v_{w}^{2}+v_{o}^{2}$ is the velocity
of wind material near the secondary, $v_{o}$ is the orbital velocity
of the secondary, and $c_{s}$ is the sound speed of the wind. For
typical velocities of $\sim$ 10--20 \kms, $\dot{M}_{a}/\dot{M}_{w}$
ranges from $\sim$ 20\% for close binaries with $a\approx$ 3 AU,
to $\lesssim$ 0.1\% for wider binaries with $a\approx$ 100 AU. Thus,
maximum accretion rates can briefly reach roughly $10^{-5}$ \msunyr\ for
close binaries with AGB star companions.

Despite the relatively low efficiency, wind accretion can add considerable
mass to the secondary. While on the RGB and AGB, the primary star
loses most of its initial mass. For primaries with initial masses
1--7 \msun, close companions can accrete as much as 1 \msun\ from
wind accretion.

Detailed AMR simulations of wind-accretion disks yield results consistent
with these simple estimates \citep{dev+09}. When the star serves
as a sink term in the simulations, circumstellar material forms a
fairly stable disk with a total mass of $\sim10^{-4}$ of the stellar
mass out to radii of roughly $5$ AU. Typically, $\sim5$\% of the
mass lost from the AGB star goes through the disk, consistent with
the simple Bondi-Hoyle estimates.

To develop a model for the structure of the disk, we assume the primary
loses mass at a constant rate, $\dot{M}_{w}$. For wide binaries,
wind material is ejected far from the accretor at approximately parallel
trajectories. Gravitational focusing close to the accretor then sets
the cross-section for material hitting the disk. Instead of the geometrical
cross-section, the fraction of the wind captured by the disk up to
distance $r$ is $f_{\dot{M}_{acc}}(<r)\propto r$. For an outer radius
equal to $R_{a}$, \begin{equation}
f_{\dot{M}_{acc}}(<R)=\frac{R}{R_{a}}.\label{eq:frac_wind}\end{equation}
 The differential accretion rate per radial bin in the disk, is therefore
\begin{equation}
\dot{M}_{acc}(R)dR=\dot{M}_{acc}^{tot}\frac{dR}{R_{a}}~.\label{eq:MdR}\end{equation}
 The source function, the instantaneous surface density profile of
material accreted into the disk is \begin{equation}
\dot{\Sigma}_{S}(R)dR=\frac{\dot{M}_{acc}(R)}{2\pi R}=\frac{\dot{M}_{acc}^{tot}dR}{2\pi RR_{a}}.\label{eq:source term}\end{equation}
 As wind material falls into the disk, it dissipates kinetic energy.
Together with irradiation from the central star and viscous heating,
this energy contributes to the heating of the disk. The amount of
energy input into an annulus at radius $r$ is the gravitational binding
energy, \begin{equation}
\dot{E}_{S}(R)dR=\frac{GM_{2}\dot{M}_{acc}^{tot}}{R}\frac{dR}{R_{a}}.\label{eq:E-accretion}\end{equation}

Although the disk does not accrete material beyond $R_{a}$, viscous
processes can expand the disk beyond $R_{a}$. However, tidal forces
from the binary companion limit the outer radius. Typically, this
outer radius is roughly 90\% to 95\% of the Roche limit, $a_{2}=r_{l}a$,
where \begin{equation}
r_{l}=\frac{0.49q^{2/3}}{0.6q^{2/3}+ln(1+q^{1/3})}~\label{eq:r_roche}\end{equation}
 and $q$ = $M_{2}/M_{1}$ \citep{egg83}.

For wind mass loss with $M_{1}>M_{2}$ and low accretion efficiency,
binaries expand. We approximate this expansion assuming adiabatic
mass loss from the primary star \citep{had63}
\begin{equation}
a(t)=\frac{M_{1,init}+M_{2,init}}{M_{1}(t)+M_{2}(t)}a_{init},\label{eq:adiabatic}\end{equation}
 where $M_{i,init}$ are the initial masses of the binary components,
$a_{init}$ is the initial separation and \begin{equation}
M_{i}(t)=M_{i,init}+\int_{0}^{t}\dot{M}_{i}(t)dt,\label{eq:mass-loss}\end{equation}
 for $i=1,2$. Assuming a constant mass loss rate from the primary,
$M_{1}(t)=M_{1,init}-\dot{M}_{w}t$ and $M_{2}(t)=M_{2,init}+\zeta\dot{M}_{w}t$,
where $\zeta$ is the accretion efficiency. For a system losing mass
in a wind ($\zeta\ll$ 1), $a(t)>a_{init}$. Thus, the binary separation expands
with time. For most mass loss rates and separations of interest, the
disk reaches a steady-state on timescales shorter than the lifetime
of the AGB star (see results in Table 1). Thus, we can safely adopt
a simple prescription for the outer radius of the disk \begin{equation}
R_{out}=a_{2}\label{eq:R_out}\end{equation}
 and assume that $a$ is constant in time. Considering a broad range
of $a$ and $\dot{M}_{w}$ allows us to derive conditions in the disk
for a broad range of plausible binaries (see Table 1).

The inner boundary of the disk, $R_{in}$, is limited by the radius
of the accreting star\begin{equation}
R_{in}=R_{2};\label{eq:R_in}\end{equation}
we adopt an inner radius of $1.5R_{2};$ the precise value of the inner 
radius has little impact on the overall structure of the wind-filled
accretion disk.

Equipped with the necessary initial and boundary conditions, we now 
discuss a model for the evolution of the disk.

\section{Disk Evolution}

\subsection{Numerical calculation}

For a disk with surface density $\Sigma$ and viscosity $\nu$, conservation
of angular momentum and energy leads to a non-linear diffusion equation
for the time evolution of $\Sigma$ \citep[e.g.,][]{lyn+74,pri81},
\begin{equation}
\frac{\partial\Sigma}{\partial t}=3R^{-1}\frac{\partial}{\partial R}\left(R^{1/2}\frac{\partial}{\partial R}\{\nu\Sigma R^{1/2}\}\right)+\left(\frac{\partial\Sigma}{\partial t}\right)_{ext}.\label{eq: disk-evol}\end{equation}
 The first term is the change in $\Sigma$ from viscous evolution;
the second term is the change in $\Sigma$ from other processes, including
mass loss from photoevaporation \citep[e.g.][]{ale+06} or planet formation
\citep[e.g.,][]{ale+09} and mass gain from wind material falling
into the disk (given by Eq. \ref{eq:source term}). Our approach neglects
the torque of the disk by the wind, which is small for a disk in a
binary system. For the wide binaries we consider ($30$ and $100$
AU separation), the wind momentum can compress the outer edges of the 
disk but does not affect its bulk structure. The viscosity is 
$\nu=\alpha c_{s}H$, where $c_{s}$ is the sound speed,
$H$ is the vertical scale height of the disk, and $\alpha$ is the
viscosity parameter. The sound speed is $c_{s}^{2}=\gamma kT_{d}/\mu m_{H}$,
where $\gamma$ is the ratio of specific heats, $k$ is Boltzmann's
constant, $T_{d}$ is the midplane temperature of the disk, $\mu$
is the mean molecular weight, and $m_{H}$ is the mass of a hydrogen
atom. The scale height of the disk is $H$ = $c_{s}\Omega^{-1}$,
where $\Omega=\sqrt{GM_{\star}/R^{3}}$ is the angular velocity.

To solve this equation numerically, we assume that the midplane temperature
is the sum of the energy generated by viscous ($T_{v})$ dissipation,
the energy from irradiation ($T_{I}$), and the energy of infalling
material from the wind ($T_{W}$), \begin{equation}
T_{d}^{4}=T_{V}^{4}+T_{I}^{4}+T_{W}^{4}.\label{eq: tdisk}\end{equation}
 The viscous temperature is \begin{equation}
T_{V}^{4}=\frac{27\kappa\nu\Sigma^{2}\Omega^{2}}{64\sigma},\label{eq: t-visc1}\end{equation}
 where $\kappa$ is the opacity and $\sigma$ is the Stefan-Boltzmann
constant \citep[e.g. ][]{rud+86,rud+91}. With $\nu=\alpha c_{s}^{2}\Omega^{-1}$
and $t_{2}=(27\alpha/64\sigma)\sim(\gamma k/\mu m_{H})\sim\kappa\Omega\Sigma^{2}$,
the viscous temperature is \begin{equation}
T_{V}^{4}=t_{2}T_{d}.\label{eq: t-visc2}\end{equation}

The energy from infalling material follows from Eq. \ref{eq:E-accretion}.
Balancing the kinetic energy of infall with the energy radiated by
the disk yields \begin{equation}
\sigma2\pi RdRT_{W}^{4}(R)=\frac{GM_{2}\dot{M}_{acc}dR}{RR_{a}}.\end{equation}
 Solving this equation for $T_{W}$ requires \begin{equation}
\sigma T_{W}^{4}=\frac{GM_{2}\dot{M}_{acc}}{2\pi R^{2}R_{a}}.\label{eq:wind-heating}\end{equation}

If the disk is vertically isothermal, the irradiation temperature
is $T_{I}^{4}(R)=(\theta/2)(R_{\star}/R)^{2}T_{\star}$, where $R_{\star}$
and $T_{\star}$ are the radius and effective temperature of the central
star and \citep{chi+97} \begin{equation}
\theta=\frac{4}{3\pi}\left(\frac{R_{\star}}{R}\right)^{3}+R\frac{\partial(H/R)}{\partial R}.\end{equation}

Thus, the irradiation temperature is \begin{equation}
\left(\frac{T_{I}}{T_{\star}}\right)^{4}=\frac{2}{3\pi}\left(\frac{R_{\star}}{R}\right)^{3}+\frac{H}{2R}\left(\frac{R_{\star}}{R}\right)^{2}\left(\frac{\partial lnH}{\partial{\rm ln}R}-1\right).\label{eq: t-irrad1}\end{equation}
 Following \citet{chi+10} and \citet{hue+05}, we set $\partial{\rm ln}H/\partial{\rm ln}R=9/7$.
With $H=c_{s}\Omega^{-1}$, we set $t_{0}=(2T_{\star}/3\pi)(R_{\star}/R)^{3}$
and $t_{1}=(R_{\star}/R)^{2}\sim(7R\Omega)^{-1}\sim(\gamma k/\mu m_{H})^{1/2}\sim T_{\star}$.
The irradiation temperature is then \begin{equation}
T_{I}^{4}=t_{0}+t_{1}T_{d}^{1/2}.\label{eq: t-irrad2}\end{equation}
 Viscous disks are not vertically isothermal \citep{rud+91,dal+98};
however this approach yields a reasonable approximation to the actual
disk structure.

Because $T_{V}$ and $T_{I}$ are functions of the midplane temperature,
and $T_{W}$ is independent of it, we solve equation (\ref{eq: tdisk})
with a Newton-Raphson technique. Using equations (\ref{eq: t-visc2}),
(\ref{eq: t-irrad2}), and \ref{eq:wind-heating}, we re-write equation
(\ref{eq: tdisk}) as \begin{equation}
f(T_{d})=T_{d}^{4}-(t_{0}+t_{1}T_{d}^{1/2}+t_{2}T_{d}+T_{W})=0.\label{eq: tdisk-nr}\end{equation}
 Adopting an initial $T_{d}\approx t_{2}^{1/3}$ or $T_{d}\approx t_{1}^{2/7}$,
the derivative \begin{equation}
\frac{\partial f}{\partial T_{d}}=4T_{d}^{3}-\frac{t}{2}T_{d}^{-1/2}-t_{2}\label{eq: deriv-nr}\end{equation}
 allows us to compute \begin{equation}
\delta T_{d}=f\left(\frac{\partial f}{\partial T_{d}}\right)^{-1}\label{eq: delta-t}\end{equation}
 and yields a converged $T_{d}$ to a part in $10^{8}$ in 2--3 iterations.

In the inner disk, the temperature is often hot enough to vaporize
dust grains. To account for the change in opacity, we follow \citet{cha09}
and assume \begin{equation}
\kappa=\kappa_{0}\left(\frac{T_{d}}{T_{e}}\right)^{n}\label{eq: kappa-evap}\end{equation}
 with $n=-14$ in regions with $T_{d}>T_{e}=$ 1380 K \citep{rud+91,ste98}.
For simplicity, we assume $\kappa=\kappa_{0}$ when $T_{d}<T_{e}$.

To solve for the time evolution of $\Sigma$, we use an explicit technique
with $N$ annuli on a grid extending from $x_{in}$ to $x_{out}$
where $x$ = 2 $R^{1/2}$ \citep{bat+81,bat+82a}, and $R_{in}$ and
$R_{out}$ are inner and outer boundaries determined by Eqs. \ref{eq:R_out}.
To verify this code, \citet{bro+11} compare numerical solutions with
analytic results from \citet{cha09}.

\subsection{Potential caveats}

\label{sub:assumptions}

Our study explores wind-accretion disks using simple numerical models.
These models provide us with the ability to explore the long term 
evolution of disks with a wide range of initial conditions. Because 
these disks reach an approximate steady-state, we can test whether 
the solutions scale as expected with the binary separation and the
mass loss rate of the primary star  However, the approach makes 
many simplifying assumptions, and does not consider many physical 
processes known to occur in these systems. Because the physics we do
not include is often poorly understood or weakly constrained, it 
seems prudent to begin with a simple physical structure and add 
additional physics as needed to understand a broader range of 
phenomena. The following paragraphs discuss useful physics to
consider for future studies.

\textbf{Radiative heating of the disk by the companion:} Our calculations
do not include radiative heating of the disk by the primary star (mass donor). 
If the primary has a radius $R_{1}\approx$ 0.5 AU and a luminosity $L_{1}$, 
a flared disk surrounding the secondary with outer radius $R_{out}$ and 
vertical scale height $h_{out}\approx$ 0.05 $R_{out}$ intercepts
0.4\% to 0.01\% of $L_{1}$ for $a$ = 3--100~AU \citep{kh1987}. With
$L_{1}\approx$ 500--1000 \lsun, this contribution to disk heating
is comparable to irradiation by the central MS star. In wide binaries
with $a\approx$ 10--100 AU, AGB stars can have $R_{1}\approx$ 1
AU. The disk then intercepts 0.2\% ($a=10$ AU) to 0.02\% ($a$ =
100 AU) of the radiation from the primary. For typical AGB $L_{1}\sim10^{4}$
\lsun\ and $a\approx$ 10 AU, this component is somewhat larger
than irradiation from the central star. In all configurations, the outer 
rim of the disk intercepts nearly all of this radiation. If the outer 
disk radiates efficiently, this extra heating should not impact the 
structure of material in the inner disk. To test this assumption, we 
performed several test simulations with a fixed outer boundary temperature,
$\sigma T_{out}^{4}=f_{irr}L_{1}/4\pi R_{out}h_{out}$. This extra
heating reduces the disk surface density by a factor of $\sim$ 2
near the outer boundary but changes the surface density at smaller
radii, $R\lesssim0.95R_{out}$, by less than 1\%. Thus, neglecting
this component has little impact on our results. If the outer disk
radiates inefficiently, the outer disk will expand and transport
thermal energy radially inward. If the scale height of this puffed
up disk becomes comparable to the accretion radius $R_{a}$ it could
significantly affect the accretion process, and would require a different
approach.

\textbf{Disk viscosity:} The processes underlying the origin of the viscosity 
in accretion disks are still not understood. Here, we follow most studies and
adopt a simple prescription using the $\alpha$ viscosity parameter. In particular, 
we do not directly consider here any interactions of magnetic fields and their 
evolution, and we use a constant $\alpha$ in each model. 
Our general results are shown our for a choice of $\alpha=0.01$. Table 1 
also provides additional data for the cases of $\alpha=0.001$ and $\alpha=0.1$,
enabling us to consider the overall dependence of disk structure and evolution
on the viscosity parameter.  

\textbf{Binary orbital expansion:} As briefly discussed above, we
do not change the binary separation and mass in our calculation. In
most cases the wind-accretion disk evolves to a steady state in a relatively 
short time scale. Our approach then provides a good approximation to reality. 
When evolution in the binary separation occurs more rapidly than the
time-scale to reach steady-state conditions in the disk (e.g., when the
time-scale to reach steady-state is $\gtrsim 10^{5}$ yr), our approximation 
is less accurate. Our results should then be taken with more caution.
In general, both the evolution of the binary and the equilibrium disk mass
evolve slowly enough to treat the time evolution with a sequence of 
steady disk models in binaries with increasing semi-major axis.

\textbf{Accretion model and disk instabilities:} We use a simplified 1D model for the 
accretion infall of wind material into the disk. Although this model captures many
of the main properties of the material infall, reality is probably more 
complex. Aside from our failure to treat the complex interaction between 
the wind and the outer disk, our simplified 1.5D model does not allow for
variations in the vertical structure which might lead to thermal instabilities
and dwarf nova outbursts \citep[e.g.,][and references therein]{cann12} or for
winds driven by radiation pressure from the inner disk \citep[e.g.,][]{noe+10}.
In particular, the inner region where the temperatures rise above 1000--2000 K, 
(roughly where dust grains evaporate) might become thermally unstable (see e.g. \citealt{ale+11}). 
Though the location of the disk instability regime depends on the opacity model,
the mass loss rate of the AGB primary, and other input parameters,
Fig. \ref{fig:Temperature-profile} shows that temperatures as high as 1000 K are 
typically achieved only for a small region, extending up to an AU for the models with 
the highest accretion rate and/or the smallest binary separations. In most models, 
these relatively high temperatures are achieved only over small regions in the inner
0.1--0.3 AU of the disk.  In models with the smallest binary separation (3 AU), the
outer disk radius is comparable to disks studied in other contexts \citep[e.g.,][]{ale+11}. 
In wider binaries with separations of 10--100 AU, the outer disk extends well beyond 
regions likely capable of sustaining a thermal instability cycle. Thus, potentially 
unstable regions can extend over a large fraction of the disk in compact binaries 
($a \approx$ 3 ~AU) but impact a fairly small fraction of the disk in wider binaries
($a \gtrsim$ 10--100~AU).  We therefore conclude that models exploring small separations, particularly those with 
the highest accretion rates (e.g. models 1 and 5 in table 1 below), could be highly 
susceptible to disk instabilities in a significant fraction of the inner disk.  We 
caution that our steady state solutions might not well represent these cases. 
The steady state solution is likely to well represent likely outcomes in wider binaries 
and in compact binaries with low accretion rates. In all cases, likely unstable regions
are close to the accreting star and could could lead to outbursts with distinct 
observational outcomes. In Alexander et al. (2011), the optical brightness changes little 
during the outbursts; the bolometric luminosity changes by a relatively small factor 
of a few for the largest infall rates, though these are larger than those we study. 

\section{Results}

To develop an understanding of the long term behavior of wind-fed
disks in wide binaries, we consider a grid of models in $(a,\dot{M}_{wind})$
space (see Tables 1 and 2 for model parameters and results). For 
simplicity, we adopt $q$ = 3, $M_{2}$ = 1 \msun\ (MS secondary) or 
$M_{2}$ = 0.6 \msun\ (WD secondary), and assume constant separation 
and mass loss rate with time. For each $(a,\dot{M}_{wind})$ pair, 
we set $v_{w}$ = 10 \kms\ and $c_{s}$ = 20 \kms, derive $R_{a}$ from 
equation (\ref{eq:r_bondi}) and $R_{out}$ from equation (\ref{eq:R_out}). 
The disk inner cut-off is $R_{{\rm in}}=1.4R_{2}$,
with $R_2$ = 1 $R_{\odot}$ for the main sequence stars and
$R_2$ = 0.01 $R_{\odot}$ for the white dwarf stars. 

For an initial disk mass $M_{d,0}$ = $10^{-7}$ \msun, the initial
surface density distribution is $\Sigma=\Sigma_{0}R^{-1}e^{-R/R_{out}}$
with $\Sigma_{0}=M_{d,0}/2\pi RR_{out}$. Using our numerical solution
to the diffusion equation, we continuously add material to the disk
inside $R_{a}$ and evolve the surface density in time until (and
if) the disk reaches a steady state at time $\tau_{steady}$, or up
to $1-2$ Myr, corresponding to the maximum lifetimes of AGB stars.
As long as the steady-state disk mass $M_{steady}$ is much larger
than the initial disk mass, numerical tests show that the initial
surface density distribution has little impact on the final steady-state
surface density distribution.

\begin{figure}
\includegraphics[scale=0.4]{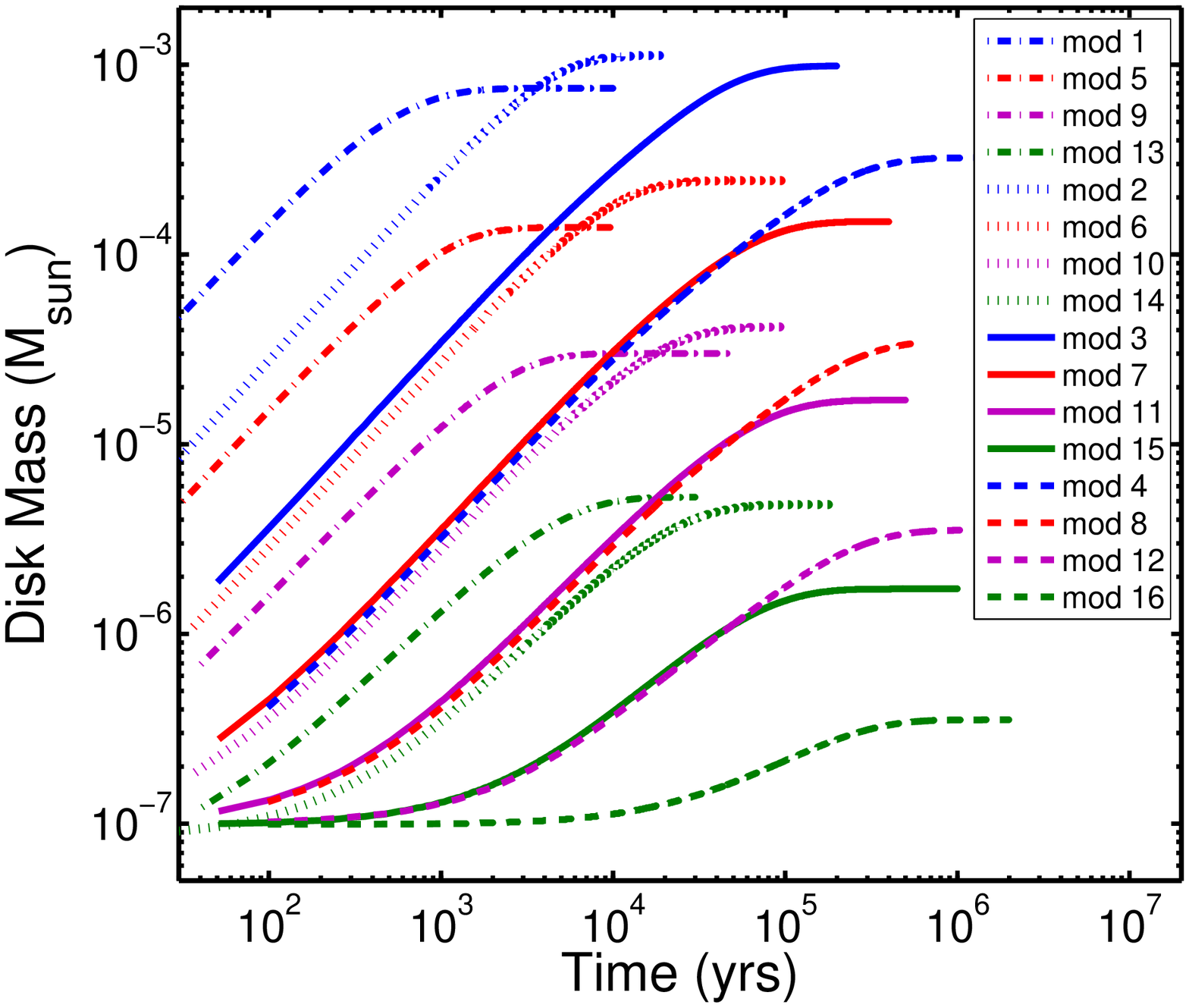}

\includegraphics[scale=0.4]{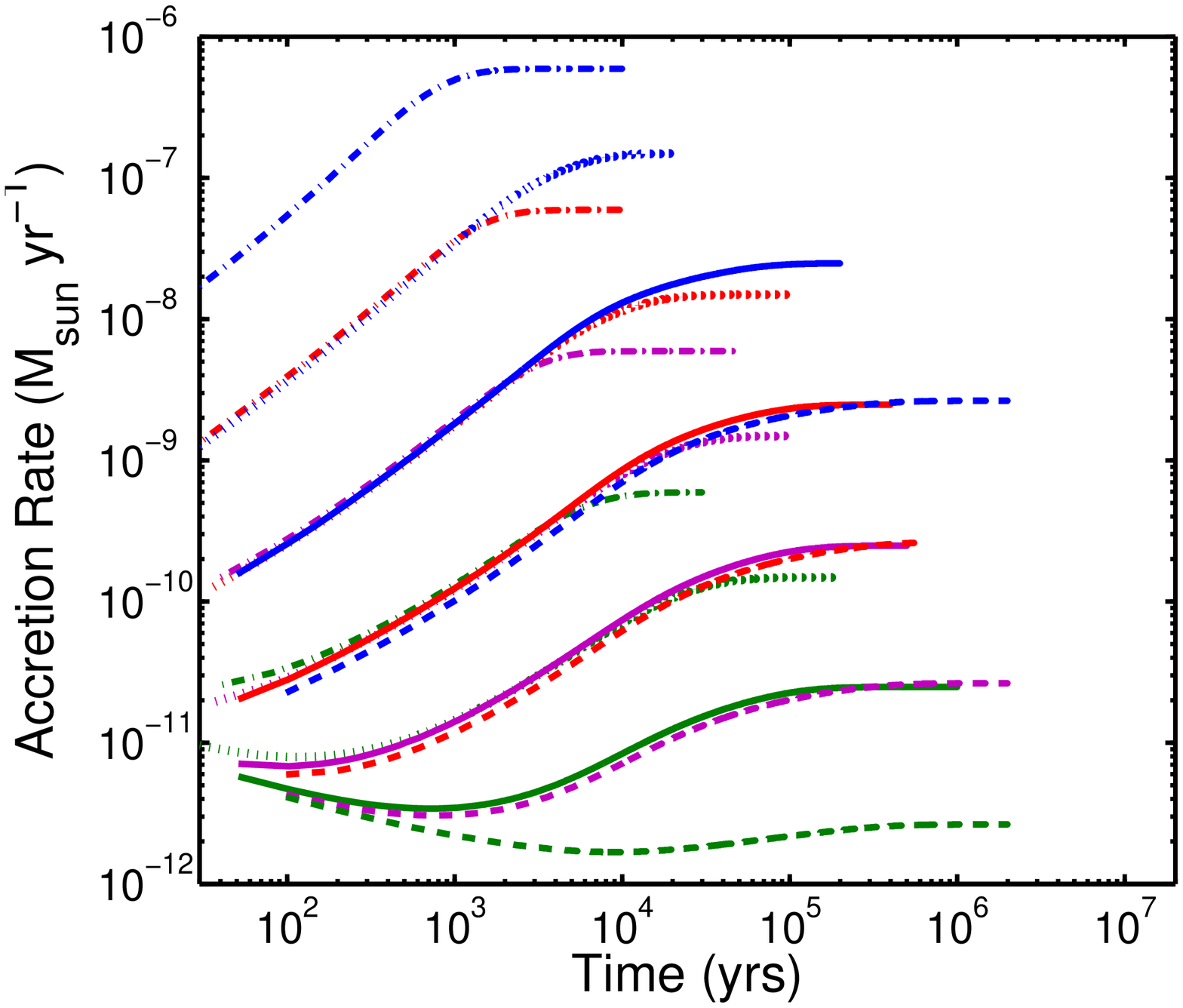}

\caption{\label{fig:mass-profile}Disk mass profile and evolution. Top: The
evolution of the total mass in the accretion disk. Disks in binaries
with larger separations are wider and accommodate more mass, as they
are truncated at further distances from the star. Bottom: Evolution
of the accretion rate onto the star. Both upper figures show how the
disk evolves into a steady state configuration over timescales which
are typically much shorter than the mass loss time scales. 
 The lines correspond to the models described in Table 1. Lines
of the same color correspond to similar mass loss from the companion
($10^{-5},$ $10^{-6},\,10^{-7}$ and $10^{-8}$ M$_{\odot}$ yr$^{-1}$
from top to bottom). Lines of the same type correspond to similar
binary separation ($3$ AU, dash-dotted; $10$ AU, dotted; $30$ AU,
solid; $100$ AU, dashed).}

\end{figure}

We consider the system to reach a steady state when the change in disk
mass per unit time is smaller than $10^{-3}$ of the input rate. In most 
cases, disks reach steady state on timescales much shorter than the AGB 
lifetimes (see Fig. \ref{fig:mass-profile}). Several models with the largest
binary separations do not properly fulfill this steady state condition
by the end of the simulation, but the actual change in the disk structure
at late times is small. For these models we show the disk properties at 
the end of the simulation run time, $\tau_{run}$. In these cases, 
$\tau_{run}$ is already comparable or longer than the typical AGB lifetime.
Thus, wind accretion disk cannot achieve a steady state configuration. 
Nevertheless the disk structure in these cases hardly changes over most 
of the AGB lifetime.

\begin{figure}
\includegraphics[scale=0.4]{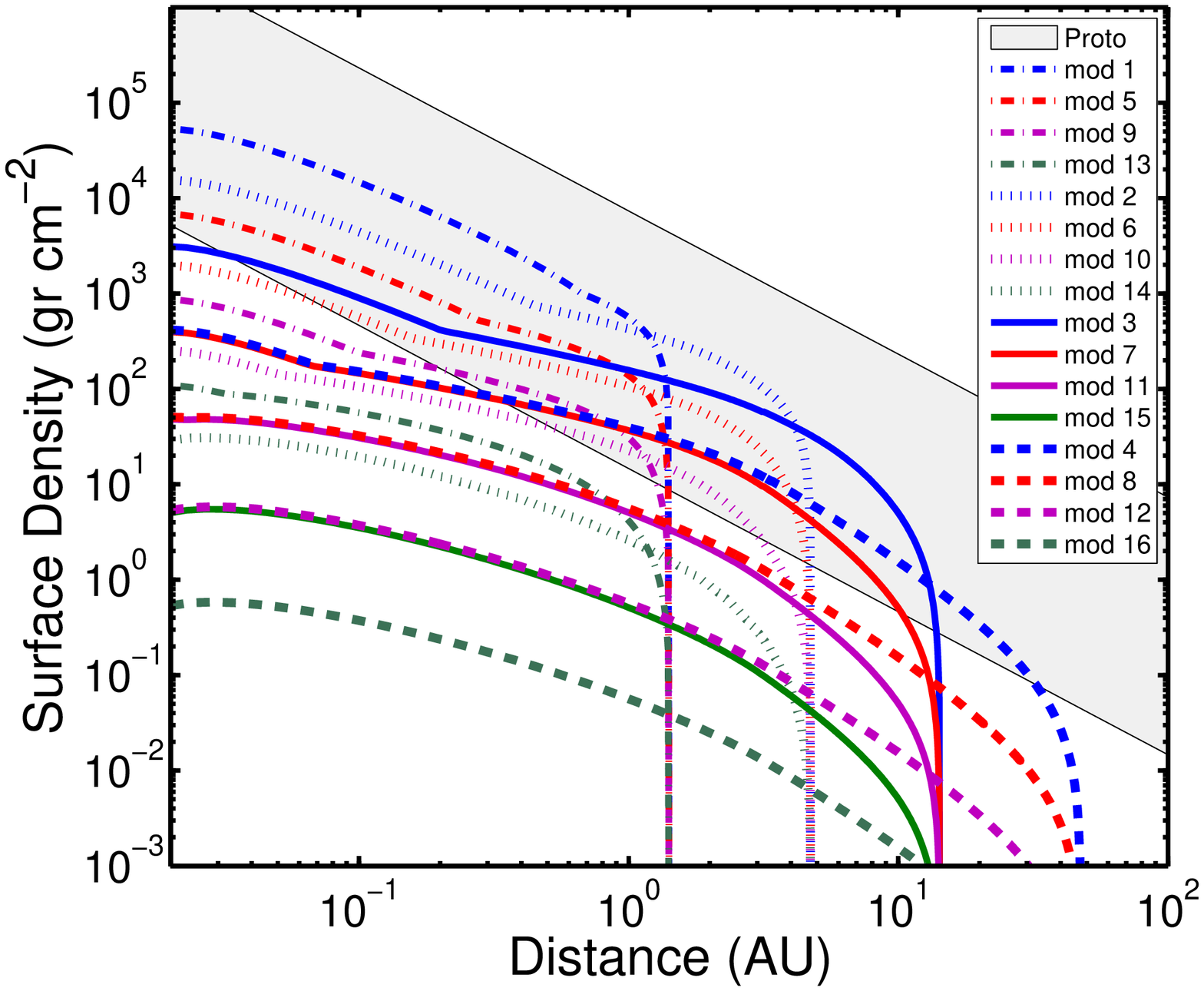}
\includegraphics[scale=0.4]{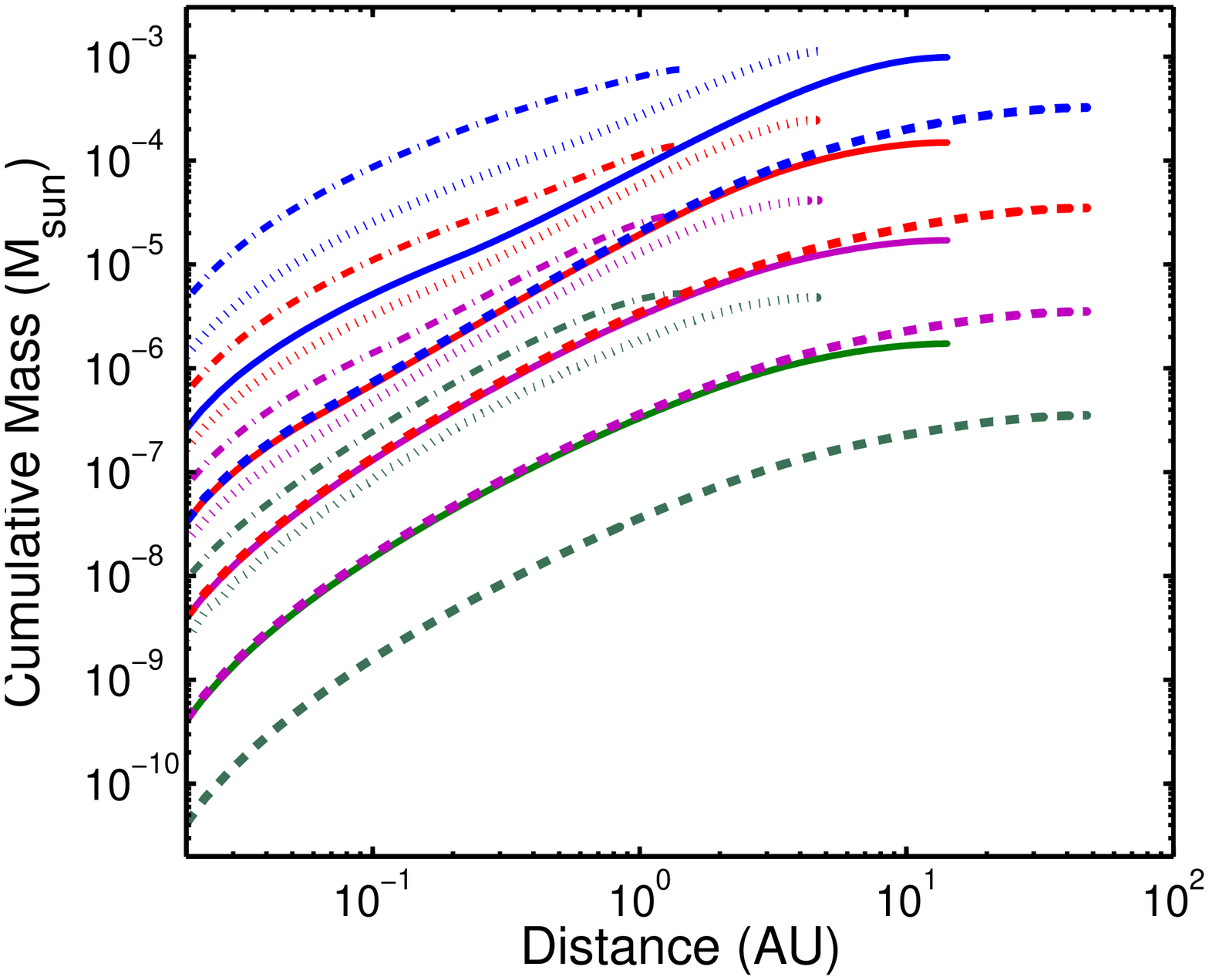}
\caption{\label{fig:Density-profile}Radial profile of wind accretion
disks at steady state. Top:  The radial density profiles of wind accretion disks under various conditions.  Also shown is a shaded region corresponding to a range
of protoplanetary disk models which only differ in their overall normalization
corresponding to the range of total disk masses as inferred from observations
($\sim2\times10^{-4}-10^{-1}$ M$_{\odot}$; mean of $2\times10^{-3}$
M$_{\odot}$; \citealp{and+07}). Bottom: The cumulative mass radial profile of wind accretion disks.  Lines correspond to the same models in Fig. 1 (described in Table 1). }

\end{figure}

\begin{figure}
\includegraphics[scale=0.4]{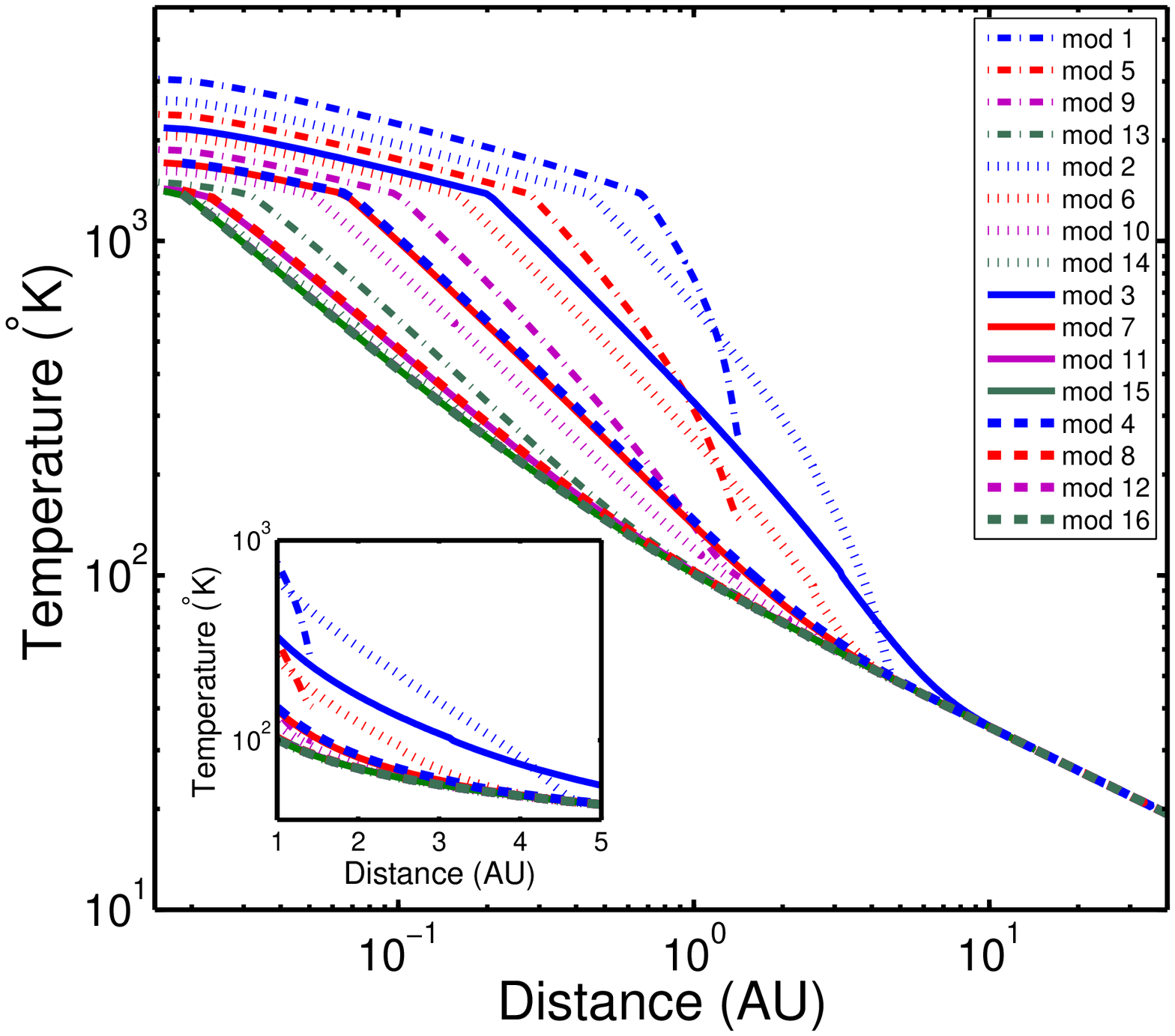}

\caption{\label{fig:Temperature-profile}Radial temperature profile of wind
accretion disks (inset shows a zoom up region; note linear distance-scale).
The radial temperature profiles of wind accretion disks under various
conditions are shown; the lines correspond to similar models as described
in Fig. 1. Note that regions in which temperatures rise beyond approximately 1000 K
might be susceptible to thermal disk instability, not modeled here.  }

\end{figure}

Tables 1--2 summarize the main results of the calculations.  The detailed 
steady state profile and evolution of the disks and the accretion rate 
(on to the star) are shown in 
Figs. \ref{fig:mass-profile}-\ref{fig:Temperature-profile}. Although
we focus on main sequence star accretors in the Figures, results for 
WD accretors are similar. Table 2 shows results for white dwarfs with
negligible luminosity. If the accreting star is a hot white dwarf with 
L = 100 $L_{\odot}$, UV photons heat up disk regions close to the 
central star but impact the outer disk very weakly. Thus, the time 
evolution of the accretion rate and disk mass (in Fig. 1) and the 
radial profiles of cumulative mass, surface density, and disk temperature 
(Figs. 2--5) are fairly representative of WD accretors and MS accretors.

\begin{table*}
\begin{raggedright}{\tiny{} \caption{MS secondary models}
}{\tiny \par}
\hspace{0.3em}
\tiny
\begin{tabular}{|l@{\hspace{0.5em}}|l@{\hspace{0.5em}}|l@{\hspace{0.5em}}|l@{\hspace{0.5em}}|l@{\hspace{0.5em}}|l@{\hspace{0.5em}}|l@{\hspace{0.5em}}|l@{\hspace{0.5em}}|l@{\hspace{0.5em}}|l@{\hspace{0.5em}}|l@{\hspace{0.5em}}l@{\hspace{0.7em}}l@{\hspace{0.5em}}|l@{\hspace{0.5em}}l@{\hspace{0.5em}}l@{\hspace{0.5em}}|l@{\hspace{0.5em}}l@{\hspace{0.5em}}l@{\hspace{0.5em}}|}

\hline 
{\tiny \#$^{1}$  } & {\tiny $m_{1}$$^{2}$} & {\tiny $m_{2}$$^{3}$} & {\tiny $a$$^{4}$} & {\tiny $R_{a}^{5}$} & {\tiny $R_{out}^{6}$} & {\tiny $\dot{M}_{loss}^{7}$} & {\tiny $\dot{M}_{disk}^{8}$} & {\tiny $M_{{\rm disk}}^{0.5}$$^{9}$} & {\tiny $\dot{M}_{in}^{10}$} & {\tiny $M_{disk}^{11}$} &  &  & {\tiny $\tau_{steady}^{12}$} &  &  & {\tiny $\tau_{run}^{13}$} &  & \tabularnewline
\hline 
 &  &  &  &  &  &  &  &  &  & {\tiny $10^{-3}$} & {\tiny $10^{-2}$} & {\tiny $10^{-1}$} & {\tiny $10^{-3}$} & {\tiny $10^{-2}$} & {\tiny $10^{-1}$} & {\tiny $10^{-3}$} & {\tiny $10^{-2}$} & {\tiny $10^{-1}$}\tabularnewline
\hline 
\hline 
{\tiny 1} & {\tiny 3.0} & {\tiny 1.0} & {\tiny 3.0} & {\tiny 1.5} & {\tiny 1.4} & {\tiny $-$5.0} & {\tiny $-$5.74} & {\tiny $-$0.05} & {\tiny $-$6.23} & {\tiny -2.20} & {\tiny $-$3.12} & {\tiny -4.01} & {\tiny 4.51} & {\tiny 3.51} & {\tiny 2.57} & {\tiny 5.00} & {\tiny 4.00} & {\tiny 3.00}\tabularnewline
{\tiny 2} & {\tiny 3.0} & {\tiny 1.0} & {\tiny 10.0} & {\tiny 2.5} & {\tiny 4.8} & {\tiny $-$5.0} & {\tiny $-$6.55} & {\tiny $-$0.85} & {\tiny $-$6.82} & {\tiny -2.12} & {\tiny $-$2.96} & {\tiny -3.77} & {\tiny 5.08} & {\tiny 4.28} & {\tiny 3.51} & {\tiny 5.30} & {\tiny 4.30} & {\tiny 3.70}\tabularnewline
{\tiny 3} & {\tiny 3.0} & {\tiny 1.0} & {\tiny 30.0} & {\tiny 3.1} & {\tiny 14.3} & {\tiny $-$5.0} & {\tiny $-$7.46} & {\tiny $-$1.74} & {\tiny $-$7.60} & {\tiny -2.14} & {\tiny $-$3.00} & {\tiny -3.89} & {\tiny 6.11} & {\tiny 5.28} & {\tiny 4.51} & {\tiny 6.30} & {\tiny 5.30} & {\tiny 4.70}\tabularnewline
{\tiny 4} & {\tiny 3.0} & {\tiny 1.0} & {\tiny 100.0} & {\tiny 3.4} & {\tiny 47.6} & {\tiny $-$5.0} & {\tiny $-$8.49} & {\tiny $-$2.80} & {\tiny $-$8.59} & {\tiny -2.53} & {\tiny $-$3.49} & {\tiny -4.47} & {\tiny 6.97} & {\tiny 6.30} & {\tiny 5.00} & {\tiny 6.97} & {\tiny 6.30} & {\tiny 5.30}\tabularnewline
{\tiny 5} & {\tiny 3.0} & {\tiny 1.0} & {\tiny 3.0} & {\tiny 1.5} & {\tiny 1.4} & {\tiny $-$6.0} & {\tiny $-$6.74} & {\tiny $-$1.05} & {\tiny $-$7.23} & {\tiny -3.00} & {\tiny $-$3.85} & {\tiny -4.70} & {\tiny 4.54} & {\tiny 3.67} & {\tiny 2.84} & {\tiny 5.00} & {\tiny 4.00} & {\tiny 3.00}\tabularnewline
{\tiny 6} & {\tiny 3.0} & {\tiny 1.0} & {\tiny 10.0} & {\tiny 2.5} & {\tiny 4.8} & {\tiny $-$6.0} & {\tiny $-$7.55} & {\tiny $-$1.85} & {\tiny $-$7.82} & {\tiny -2.77} & {\tiny $-$3.62} & {\tiny -4.49} & {\tiny 5.51} & {\tiny 4.66} & {\tiny 3.83} & {\tiny 6.00} & {\tiny 5.00} & {\tiny 4.00}\tabularnewline
{\tiny 7} & {\tiny 3.0} & {\tiny 1.0} & {\tiny 30.0} & {\tiny 3.1} & {\tiny 14.3} & {\tiny $-$6.0} & {\tiny $-$8.46} & {\tiny $-$2.74} & {\tiny $-$8.60} & {\tiny -2.89} & {\tiny $-$3.82} & {\tiny -4.80} & {\tiny 6.51} & {\tiny 5.51} & {\tiny 4.52} & {\tiny 6.60} & {\tiny 5.60} & {\tiny 4.60}\tabularnewline
{\tiny 8} & {\tiny 3.0} & {\tiny 1.0} & {\tiny 100.0} & {\tiny 3.4} & {\tiny 47.6} & {\tiny $-$6.0} & {\tiny $-$9.49} & {\tiny $-$3.80} & {\tiny $-$9.59} & {\tiny -3.47} & {\tiny $-$4.46} & {\tiny -5.46} & {\tiny 7.00} & {\tiny 6.00} & {\tiny 5.04} & {\tiny 7.23} & {\tiny 6.00} & {\tiny 5.30}\tabularnewline
{\tiny 9} & {\tiny 3.0} & {\tiny 1.0} & {\tiny 3.0} & {\tiny 1.5} & {\tiny 1.4} & {\tiny $-$7.0} & {\tiny $-$7.74} & {\tiny $-$2.05} & {\tiny $-$8.23} & {\tiny -3.68} & {\tiny $-$4.52} & {\tiny -5.40} & {\tiny 5.51} & {\tiny 4.51} & {\tiny 3.51} & {\tiny 5.70} & {\tiny 4.70} & {\tiny 3.70}\tabularnewline
{\tiny 10} & {\tiny 3.0} & {\tiny 1.0} & {\tiny 10.0} & {\tiny 2.5} & {\tiny 4.8} & {\tiny $-$7.0} & {\tiny $-$8.55} & {\tiny $-$2.85} & {\tiny $-$8.82} & {\tiny -3.47} & {\tiny $-$4.39} & {\tiny -5.34} & {\tiny 5.83} & {\tiny 4.96} & {\tiny 4.04} & {\tiny 6.00} & {\tiny 5.00} & {\tiny 4.30}\tabularnewline
{\tiny 11} & {\tiny 3.0} & {\tiny 1.0} & {\tiny 30.0} & {\tiny 3.1} & {\tiny 14.3} & {\tiny $-$7.0} & {\tiny $-$9.46} & {\tiny $-$3.74} & {\tiny $-$9.60} & {\tiny -3.80} & {\tiny $-$4.77} & {\tiny -5.77} & {\tiny 6.52} & {\tiny 5.54} & {\tiny 4.54} & {\tiny 6.70} & {\tiny 5.70} & {\tiny 4.70}\tabularnewline
{\tiny 12} & {\tiny 3.0} & {\tiny 1.0} & {\tiny 100.0} & {\tiny 3.4} & {\tiny 47.6} & {\tiny $-$7.0} & {\tiny $-$10.49} & {\tiny $-$4.80} & {\tiny $-$10.59} & {\tiny -4.46} & {\tiny $-$5.46} & {\tiny -6.46} & {\tiny 7.04} & {\tiny 6.04} & {\tiny 5.00} & {\tiny 7.30} & {\tiny 6.30} & {\tiny 5.30}\tabularnewline
{\tiny 13} & {\tiny 3.0} & {\tiny 1.0} & {\tiny 3.0} & {\tiny 1.5} & {\tiny 1.4} & {\tiny $-$8.0} & {\tiny $-$8.74} & {\tiny $-$3.05} & {\tiny $-$9.23} & {\tiny -4.37} & {\tiny $-$5.28} & {\tiny -6.24} & {\tiny 5.23} & {\tiny 4.36} & {\tiny 3.43} & {\tiny 5.48} & {\tiny 4.48} & {\tiny 3.48}\tabularnewline
{\tiny 14} & {\tiny 3.0} & {\tiny 1.0} & {\tiny 10.0} & {\tiny 2.5} & {\tiny 4.8} & {\tiny $-$8.0} & {\tiny $-$9.55} & {\tiny $-$3.85} & {\tiny $-$9.82} & {\tiny -4.34} & {\tiny $-$5.32} & {\tiny -6.32} & {\tiny 6.04} & {\tiny 5.04} & {\tiny 4.04} & {\tiny 6.30} & {\tiny 5.30} & {\tiny 4.30}\tabularnewline
{\tiny 15} & {\tiny 3.0} & {\tiny 1.0} & {\tiny 30.0} & {\tiny 3.1} & {\tiny 14.3} & {\tiny $-$8.0} & {\tiny $-$10.46} & {\tiny $-$4.74} & {\tiny $-$10.60} & {\tiny -4.77} & {\tiny $-$5.77} & {\tiny -6.77} & {\tiny 6.54} & {\tiny 5.54} & {\tiny 4.51} & {\tiny 7.00} & {\tiny 6.00} & {\tiny 5.00}\tabularnewline
{\tiny 16} & {\tiny 3.0} & {\tiny 1.0} & {\tiny 100.0} & {\tiny 3.4} & {\tiny 47.6} & {\tiny $-$8.0} & {\tiny $-$11.49} & {\tiny $-$5.80} & {\tiny $-$11.59} & {\tiny -5.46} & {\tiny $-$6.46} & {\tiny -7.46} & {\tiny 7.04} & {\tiny 6.00} & {\tiny 5.04} & {\tiny 7.30} & {\tiny 6.30} & {\tiny 5.30}\tabularnewline
\hline 
\end{tabular}{\tiny \par}

\end{raggedright}

Columns: (1) Model number; (2) Primary mass ($M_{\odot}$) ; (3) Secondary
mass ($M_{\odot}$); (4) SMA separation (AU); (5) Bondi-Hoyle accretion
radius (AU) ; (6) Outer boundary of the disk (AU); (7) Mass loss rate
from primary ($M_{\odot}$ yr$^{-1}$) ; (8) Accretion rate into disk
($M_{\odot}$ yr$^{-1}$) ; (9) Total mass that goes through the disk
in 0.5 Myrs ($M_{\odot}$); (10) Accretion rate onto the star ($M_{\odot}$
yr$^{-1}$) ; (11) Disk mass at steady state ($M_{\odot}$); (12)
Time to achieve steady state (yrs); (13) Simulation run time (yrs).\\
 Columns 7-13 are shown in logarithmic scale.\label{Tab:ms} 
\end{table*}

%\par\end{flushleft}

%\begin{flushleft}
%
\begin{table*}

\begin{raggedright}
\caption{WD secondary models}
\begin{tabular}{|l|l|l|l|l|l|l|l|l|l|l|l|l|}
\hline 
\#$^{1}$ & $m_{1}$$^{2}$ & $m_{2}$$^{3}$ & $a$$^{4}$ & $R_{a}$$^{5}$ & $R_{out}$$^{6}$ & $\dot{M}_{loss}$$^{7}$ & $\dot{M}_{disk}$$^{8}$ & $M_{{\rm disk}}^{0.5}$$^{9}$ & $\dot{M}_{in}$$^{10}$ & $M_{disk}$$^{11}$ & $\tau_{steady}$$^{12}$ & $\tau_{run}$$^{13}$\tabularnewline
\hline
\hline 
1 & 1.8 & 0.6 &   3.0 & 1.2 &  1.4 &  $-$5.0 &  $-$6.06 &  $-$0.37 &  $-$6.47 &  $-$3.29 &  4.00 &  4.48 \\
2 & 1.8 & 0.6 &  10.0 & 1.7 &  4.8 &  $-$5.0 &  $-$6.96 &  $-$1.26 &  $-$7.18 &  $-$3.24 &  4.51 &  4.70 \\
3 & 1.8 & 0.6 &  30.0 & 2.0 & 14.3 &  $-$5.0 &  $-$7.89 &  $-$2.19 &  $-$8.02 &  $-$3.25 &  5.51 &  5.70 \\
4 & 1.8 & 0.6 & 100.0 & 2.1 & 47.6 &  $-$5.0 &  $-$8.96 &  $-$3.24 &  $-$9.00 &  $-$3.51 &  6.30 &  6.30 \\
5 & 1.8 & 0.6 &   3.0 & 1.2 &  1.4 &  $-$6.0 &  $-$7.06 &  $-$1.37 &  $-$7.47 &  $-$4.00 &  3.75 &  4.00 \\
6 & 1.8 & 0.6 &  10.0 & 1.7 &  4.8 &  $-$6.0 &  $-$7.96 &  $-$2.26 &  $-$8.18 &  $-$3.85 &  4.80 &  5.00 \\
7 & 1.8 & 0.6 &  30.0 & 2.0 & 14.3 &  $-$6.0 &  $-$8.89 &  $-$3.19 &  $-$9.02 &  $-$3.92 &  5.88 &  6.00 \\
8 & 1.8 & 0.6 & 100.0 & 2.1 & 47.6 &  $-$6.0 &  $-$9.96 &  $-$4.24 & $-$10.01 &  $-$4.39 &  6.30 &  6.30 \\
9 & 1.8 & 0.6 &   3.0 & 1.2 &  1.4 &  $-$7.0 &  $-$8.06 &  $-$2.37 &  $-$8.47 &  $-$4.64 &  4.51 &  4.70 \\
10 & 1.8 & 0.6 &  10.0 & 1.7 &  4.8 &  $-$7.0 &  $-$8.96 &  $-$3.26 &  $-$9.18 &  $-$4.48 &  5.51 &  5.60 \\
11 & 1.8 & 0.6 &  30.0 & 2.0 & 14.3 &  $-$7.0 &  $-$9.89 &  $-$4.19 & $-$10.02 &  $-$4.70 &  6.00 &  6.00 \\
12 & 1.8 & 0.6 & 100.0 & 2.1 & 47.6 &  $-$7.0 & $-$10.96 &  $-$5.24 & $-$11.01 &  $-$5.33 &  6.30 &  6.30 \\
13 & 1.8 & 0.6 &   3.0 & 1.2 &  1.4 &  $-$8.0 &  $-$9.06 &  $-$3.37 &  $-$9.47 &  $-$5.28 &  4.51 &  5.00 \\
14 & 1.8 & 0.6 &  10.0 & 1.7 &  4.8 &  $-$8.0 &  $-$9.96 &  $-$4.26 & $-$10.18 &  $-$5.21 &  5.54 &  5.60 \\
15 & 1.8 & 0.6 &  30.0 & 2.0 & 14.3 &  $-$8.0 & $-$10.89 &  $-$5.19 & $-$11.02 &  $-$5.60 &  6.00 &  6.00 \\
16 & 1.8 & 0.6 & 100.0 & 2.1 & 47.6 &  $-$8.0 & $-$11.96 &  $-$6.24 & $-$12.01 &  $-$6.29 &  6.30 &  6.30 \\
\hline
\end{tabular}
\par\end{raggedright}
Columns: (1) Model number; (2) Primary mass ($M_{\odot}$) ; (3) Secondary mass ($M_{\odot}$); (4) SMA separation (AU); (5) Bondi-Hoyle accretion radius (AU) ; (6) Outer boundary of the disk (AU); (7) Mass loss rate from primary ($M_{\odot}$ yr$^{-1}$) ; (8) Accretion rate into
disk ($M_{\odot}$ yr$^{-1}$) ; (9) Total mass that goes through the disk in 0.5 Myrs  ($M_{\odot}$); (10)
Accretion rate onto the star ($M_{\odot}$ yr$^{-1}$) ; (11) Disk mass at steady state ($M_{\odot}$); (12)
Time to achieve steady state (yrs); (13) Simulation run time (yrs).\\
Columns 7-13 are shown in logarithmic scale. \label{Tab:WD}
\end{table*}

\par
%\end{flushleft}

%

For all calculations, the disk mass and the accretion rate onto the star 
increase roughly linearly with time and then reach an approximate steady 
state (Figure \ref{fig:mass-profile}). The timescale to reach steady-state
increases roughly linearly with increasing binary separation. 
Although the relation is shallower than a linear one, binaries with smaller
mass infall rates also take longer to reach steady state than binaries with
large mass infall rates.  The steady state disk masses range 
from $\sim10^{-3}$~\msun\ for binaries with massive winds and small separations 
to $\sim10^{-7}$~\msun\ for binaries with low mass loss rates and large 
separations. For models with identical mass loss rates from the evolved 
companion, wider binaries have smaller infall rates into the disk and smaller 
disk masses. When binaries with different $a$ have similar infall rates, 
wider binaries have larger and more massive disks.

For most main sequence star models, wind-fed accretion rates produce 
negligible luminosities. With typical steady accretion rates smaller than 
$10^{-7}$~\msunyr, the accretion luminosity is comparable to or smaller
than the stellar luminosity of 1 \lsun. Thus, the accretion disk and
boundary layer are invisible \citep{ken+84}. For close binaries with AGB-type
primary stars, accretion rates of $10^{-6}$~\msunyr\ produce modest accretion
luminosities of $\sim$ 10~\lsun. Although optical radiation from the disk
and boundary layer are not detectable, ultraviolet (UV) radiation from the
boundary layer is probably visible \citep{ken+84}.

Models of wind-fed white dwarf stars are more interesting. In these systems, 
accretion rates exceeding $10^{-9}$~\msunyr\ yield bright UV sources which
can ionize the wind from the primary \citep{ken+84}. This ionized wind
produces bright optical and UV emission lines \citep{ken+84} and a luminous
radio sources at cm wavelengths \citep{taysea84}. These systems would be 
classified as symbiotic stars \citep{ken86}.

Despite their much lower accretion rates, the central star accretes a 
larger fraction of infalling material in wide binaries. For $a\approx$
3--100 AU, the Bondi-Hoyle accretion radius $R_{a}$ exceeds $R_{st}$
the stagnation radius where the flow velocity in the disk changes
sign. Thus, some fraction of infalling material flows out through
the disk. The rest accretes onto the central star. In these simulations,
the ratio of accreted to ejected material ranges from roughly 55\%
for close binaries ($a$ = 3 AU) to $\sim$80\% for wide binaries
($a$ = 100 AU). 
The accretion flow in the accretion disk is bimodal. Generally, material 
at separations somewhat shorter than $R_A$ flows into the star; the rest 
of the material outflows out into the extended disk up to $R_{out}$ and 
is eventually ejected from the disk.

Accreted material typically has a negligible impact on the central MS star. 
For typical lifetimes of $10^{5}$ yr or less, the central star accretes less 
than 1\% of its initial mass. In the closest binaries with the highest mass 
loss rates, the accretor may accrete from 10\% to 90\% of its initial mass
over the lifetime of the companion (e.g. see the overall accreted mass 
assuming steady state, over $0.5$ Myr, $\dot{M}_{disk}^{\,0.5}$;
Table 1). Although our models do not account for the mass increase of the 
secondary in these cases, the disk achieves steady-state long before the 
MS star accretes a significant amount of mass. Thus, our steady-state
disks still correctly represent the likely physical state of the system.

Accreted material has a more significant impact on accreting white dwarfs.
Aside from their visibility as symbiotic stars, wind-fed disks surrounding
white dwarfs can produce a variety of eruptive phenomena \citep{ken86}. 
At low accretion rates, cold white dwarfs undergo degenerate shell flashes 
and become classical novae \citep{ken+83}. White dwarfs accreting at larger
rates yield non-degenerate flashes as in the symbiotic stars V1016 Cyg and
HM Sge \citep{miko+92}. In some systems, the rates are sufficient to allow 
the white dwarf to evolve close to the Chandrasekhar limit and become a 
type Ia supernova \citep{starr+05}.

Steady-state surface density profiles generally follow expectation. Over
the entire disk, the surface density approximately scales with radius as
$\Sigma \propto r^{-n}$, with $n \approx$ 0.8--1.5.  The normalization of 
the surface density scales almost linearly with the accretion rate into 
the disk.  Changes in slope typically occur in the inner disk, when the disk 
becomes hot enough to evaporate dust grains, and in the outer disk, when the 
energy generated by irradiation or infall dominates the energy generated by 
viscosity \citep[e.g.,][and references therein]{cha09}. 

We briefly explored the effect of different viscosity parameters on the overall 
structure and mass of the disks (see table 1). In general, the steady state disk 
mass is roughly inversely proportional to $\alpha$, and scales linearly with 
$\alpha^{-1}$. These results hold as long as the disk is cold. As the disk gets 
hotter, more of the disk has an opacity dependent temperature which slightly 
changes the relation, but the overall behavior is close to linear.  Beside the 
different overall scaling, the disk spatial structure is very similar.
 
Steady-state temperature profiles are generally shallower than steady-steady
accretion disks fed by a lobe-filling companion \citep{lyn+74}. In most
steady disks, the effective temperature of the disk scales with radius as
$T \propto r^{-m}$, with $m \approx$ 0.75. For wind-fed disks in very wide 
binaries at very low accretion rates, the steady-state slope approaches the 
standard $m \approx$ 0.75. As the binary separation contracts and the wind
accretion rate grows, the temperature profile becomes more shallow. For the
closest binaries, our results suggest $m \approx$ 0.4. The range in the slope 
is small, from $m \approx$ 0.4 in close binaries with large accretion rates 
to $m \approx$ 0.7 in wide binaries with small accretion rates. Thus, the 
temperature profile scales very weakly with the separation or the mass loss
rate of the primary star. 

\section{Discussion}

The results from the calculations lead to several broad conclusions. Physical
properties of wind-fed accretion disks depend on the orbital separation and
the evolutionary state (mass loss rate) of the primary star. When the secondary
is a main sequence star, it is observable only when the binary is close and the 
primary is an AGB star. Otherwise, wind-fed accretion has few, if any, direct 
observational consequences. When the secondary is a white dwarf, the binary is
almost always observable as a symbiotic star.

Although optical/UV spectroscopic observations are sufficient to detect 
luminous symbiotic stars \citep{ken86}, direct imaging can often reveal 
the binary companion. In Mira B, IUE spectroscopic observations first 
identified emission from a hot white dwarf accreting material from the
Mira wind \citep{reim+85}, recent HST observations resolved the system
and detected the accretion disk directly \citep{ire+07}.  For a derived
binary separation of $\sim$ 70 AU, the inferred accretion luminosity of
$\sim10^{-10}\, M_{\odot}$yr$^{-1}$ \citep{sok+10} is roughly consistent 
with our models.  High resolution imaging of other symbiotic-like binaries,
including R Aqr and CH Cyg, would provide additional tests of these 
calculations.

Aside from producing interesting symbiotic stars, WD accretors in wide binaries 
can accrete enough material to approach the Chandrasekhar limit 
\citep[Table 2;][and references therein]{starr+05,she+07}. If these WDs grow,
they are susceptible to traditional type Ia supernovae \citep{arn+96} and to
sub-Chandrasekhar Helium detonation supernovae (e.g.
\citep{woo+86,bil+07,per+10,wal+11}. In most simulations, the growth of the
white dwarf depends on the way material is accreted. Thus, understanding
the structure of the disk is important for understanding whether wind-fed
white dwarfs are good SNe progenitors. Our disk models provide a step along
the path to understanding these outcomes.

Wind-fed disks around white dwarfs provide a less traditional connection to
other astrophysical systems.  For separations of $3-100$ AU and mass loss 
rates of $10^{-8}-10^{-5}$ M$_{\odot}^{-1}$, wind-fed disks have surface 
density and temperature profiles similar to those observed in low-mass 
protoplanetary disks. For comparison, Fig. \ref{fig:Density-profile} shows 
the range of surface density profiles inferred for an ensemble of protoplanetary 
disks in nearby star-forming regions \citep{and+07,chi+10}.  The shaded region 
corresponds to disks with masses of $\sim2\times10^{-4}-10^{-1}$ M$_{\odot}$. 
With typical lifetimes of a few Myr, physical processes in protoplanetary 
disks yield planets on short timescales. Given the similar lifetimes of AGB 
mass-losing primary stars, it is plausible that some aspects of planet 
formation occur in the wind-fed disks of wide binary systems \citep{per10b,per10a}. 

Despite the similarity in instantaneous disk masses, disks in wide binaries 
have a distinct advantage over protoplanetary disks in producing planets.
Continuous feeding by the primary star guarantees that the total mass available 
for planet formation is, in principle, larger for wide binaries than for 
protoplanetary disks. In close binaries with evolved AGB primaries, the
total mass accreted by the central star is at least as large as the most
protoplanetary disk. Thus, massive planets could grow in many of these disks.

Testing this idea is challenging. The most massive disks in wide binaries are 
not as large as those in protoplanetary disks; dust emission from the AGB 
primary complicates acquiring the submm and mm observations required to estimate
dust masses. Occasional nova eruptions may also frustrate the coagulation
processes that grow dust grains from the Mira wind into the planetesimals that
produce planets. However, ALMA has sufficient resolution to detect structures
in the disks of wide binaries. Identifying structures similar to those in
protoplanetary disks might allow robust comparisons between the grain properties
and the profiles of surface density and temperature.  Detection of debris disks 
around nearby white dwarfs suggests that planets (or debris from planets) survive 
the evolution from a main sequence star into a white dwarf \citep[e.g.,][]{kilic+12}.  
The frequency of wide WD companions to WD debris disks provides some estimate for 
the likelihood of this `second generation' planet formation scenario. 

\section{Summary}

Using a set of numerical calculations, we have explored the evolution and long 
term steady state structure of wind-accretion disks in wide evolved binaries 
($a \approx 3-100$ AU). These systems evolve rapidly (see Fig. \ref{fig:mass-profile}) 
and achieve a steady state on timescales which are typically much shorter than 
typical stellar evolution timescales of $\sim 10^{5}$ yr. During steady state,
disks have similar surface density and temperature profiles with total masses
of a few$\times10^{-5}-10^{-3}$ M$_{\odot}$.  The radial density profile is 
a broken power law in the inner regions and an exponential decline at the 
outer edge where the disk is truncated by tidal forces from the companion. 
The radial temperature profile is also described by a broken power law but 
does not decline dramatically at the outer edge of the disk. 

Understanding the formation, structure, and evolution of wind-fed accretion disks 
is important for a wide variety of phenomena in evolved wide binary systems, 
including chemically peculiar stars, novae, supernovae, stellar outbursts, and 
symbiotic binaries.  In close binary systems with evolved AGB primaries, significant 
accretion from a wind-fed disk provides a natural mechanism for chemical enrichment 
and abundance anomalies in otherwise normal main sequence and white dwarf stars.  
Over a broad range of separations and mass loss rates, wind-fed accretion onto 
white dwarfs can produce nova eruptions and, possibly, supernova eruptions.  
By providing a foundation for relating the mass loss rate of evolved red giants 
to the temperature and luminosity of the companion, our results also enable more 
detailed studies of the physical structure of individual binary systems.  

Our analysis begins to make a link between the disks in evolved binaries and 
protoplanetary disks. The structure and evolution of both types of disks depends
on uncertain physics, including the disk viscosity and interactions between 
stellar photons and disk material. For most accretion rates, large disks often
drive massive winds which may interact with the surrounding wind from the
primary (in a binary system) or a molecular cloud (in a protostellar system). 
Observational comparisons among disks with similar accretion rates and sizes
might improve our overall understanding of accretion phenomena.

\bibliographystyle{apj}
%\bibliography{planet-formation}

\end{document}